\documentclass[aps,prd,groupedaddress,]{revtex4}
\usepackage{latexsym}
\usepackage[dvips]{graphicx}

\newcommand{\eq}{\begin{equation}}
\newcommand{\feq}{\end{equation}}
\newcommand{\eqn}{\begin{eqnarray}}
\newcommand{\feqn}{\end{eqnarray}}
\newcommand{\arr}{\begin{eqnarray*}}
\newcommand{\farr}{\end{eqnarray*}}
\newcommand{\beq}{\begin{equation}}
\newcommand{\eeq}{\end{equation}}
\newcommand{\bea}{\begin{eqnarray}}
\newcommand{\eea}{\end{eqnarray}}

\def\beq{\begin{equation}}
\def\eeq{\end{equation}}
\def\feq{\end{equation}}
\def\bea{\begin{eqnarray}}
\def\eea{\end{eqnarray}}
\def\bc{\begin{displaymath}}
\def\ec{\end{displaymath}}
\def\lb{\label}

\def\la{\lambda}

\def\lb{\label}

\begin{document}


\title{Acoustic analogs of two-dimensional black holes    }

\author{Mariano Cadoni}
\email{mariano.cadoni@ca.infn.it}

\affiliation{Dipartimento di Fisica,
Universit\`a di Cagliari, and INFN sezione di Cagliari, Cittadella
Universitaria 09042 Monserrato, ITALY}


\begin{abstract}
We present a general method for  constructing acoustic analogs of the black 
hole solutions  of two-dimensional (2D) dilaton gravity. Because 
by dimensional reduction every 
spherically symmetric,  four-dimensional (4D) black hole admits a 2D 
description, the method can be also  used to construct analogue 
models of 4D black holes.  We also show that after fixing the gauge 
degrees of freedom the 2D gravitational 
dynamics is equivalent to an one-dimensional fluid dynamics.
 This enables  us
to find a natural definition of mass $M$, temperature $T$ and 
entropy $S$ of the 
acoustic black hole. In particular the first principle of 
thermodynamics $dM=TdS$ becomes a consequence of the fluid dynamics 
equations. 
We also discuss the general solutions 
 of the  fluid dynamics 
and two particular cases,  the 2D 
Anti-de sitter black hole and the 4D Schwarzschild black hole.

\end{abstract}


\maketitle

\section{Introduction}
In recent years it has become increasingly clear  that 
condensed matter systems can be used to mimic various kinematical 
aspects of general relativity 
\cite{Unruh:1980cg,Visser:1993ub,Visser:1997ux,Novello:2002qg,Visser:2001fe}. 
Analogue models of gravity have been used not only  to mimic 
black holes and event horizons but also   to 
describe  other spacetime structures,  cosmological solutions and field 
theory in curved spacetimes 
\cite{Barcelo:2001ca,fedichev0303,fedichev0304,
Barcelo:2003et,Barcelo:2004wz,Balbinot:2004da,Balbinot:2004dc,Berti:2004ju,
Cardoso:2004fi}.
Condensed matter analogs of gravity  could be also used to shed 
light on  old and new puzzling features of the gravitational 
interaction such as the information loss problem in the Hawking 
evaporation of a black hole or the holographic principle.
Moreover, analogue models of gravity that use condensed matter systems 
have also the nice feature of being, at least in principle, 
experimentally testable in laboratory. In the 
near future this could open the way to a ``black hole  phenomenology'' 
based not on astrophysical but   condensed 
matter experiments.

One strong limitation of this approach is that until now  analogue 
models of  black holes can be used to mimic the kinematical but not the 
dynamical aspects of  gravitational systems. For instance, one can 
use acoustic analogs to study  the kinematics of event horizons, or 
even  the Hawking radiation but not the formation of horizons due to 
some distribution of matter or the back-reaction of the geometry on the
Hawking radiation (some progress along this direction has been 
achieved in Ref. \cite{Balbinot:2004da,Balbinot:2004dc}) . 
A further unpleasant consequence of this limitation  is the 
impossibility to reproduce black hole thermodynamics using analogue 
models of black holes. It is well known that the laws of black hole 
thermodynamics are  a consequence of the Einstein equations. 
Although one can 
 define the temperature associated with  an acoustic horizon 
\cite{Visser:1997ux}, it is extremely problematic to find the right definition of 
mass and 
entropy of an acoustic black hole.

At first sight to mimic the dynamical aspects of general relativity 
using a condensed matter system seems an impossible task. The 
description of the gravitational system is characterized by a huge 
redundancy  of gauge degrees of freedom and by a separation between 
matter sources and  gravitational field (respectively described by the 
stress-energy tensor and the curvature tensors in the Einstein equations).
Conversely, focusing on acoustic black holes,  the fluid dynamics is 
characterized by few physical degrees of freedom (pressure, velocity, density) 
and does 
not seem to allow for a  source-field description .

Nonetheless, there is a situation in which one could hope to merge 
gravitational and fluid dynamics. A spherically symmetric, 
black hole with gauge 
degrees of freedom completely fixed is characterized, owing to 
Birkhoff's theorem,  by a handful of 
physical observables (the charges associated with global symmetries ). 
These observables could be put in correspondence 
with the  fluid parameters.
A spherically symmetric black hole can be described by an 
effective two-dimensional(2D) dilaton gravity model 
\cite{Grumiller:2002nm} obtained from four-dimensional(4D) Einstein gravity 
by retaining only the radial modes. 
Thus, the most promising framework for trying to merge gravitational 
and fluid dynamics is that of  2D gravity. 

In this paper we will show that this is possible.
We propose a general method for  constructing acoustic analogs of the black 
hole solutions  of two-dimensional (2D) dilaton gravity (Section II). 
Further, we  show that after fixing the gauge 
degrees of freedom the 2D gravitational 
dynamics is equivalent to a constrained one-dimensional
fluid dynamics (Section III). 
This enables  us
to find a natural definition of mass $M$, temperature $T$ and 
entropy $S$ of the 
acoustic black hole. In particular the first principle of 
thermodynamics $dM=TdS$ becomes a consequence of the fluid dynamics 
equations (Sections IV and V).
The constraint for the fluid  dynamics can be enforced in two 
independent ways. If 
 the external parameters for the fluid (profile of the flux tube or 
potential for the external forces) are given   the constraint simple 
determines the barotropic equation of state for the fluid. 
Alternatively, if one 
chooses a given Equation of state  the constraint determines the form 
of the external parameters for the fluid  (Section VI).
We also discuss  two particular cases,  the 2D 
Anti-de sitter black hole and the 4D
Schwarzschild black hole
(Section VII).

\section{2D acoustic metric and 2D black holes}

Let us consider the  generic 2D  dilaton gravity model (For a review 
see Ref. \cite{Grumiller:2002nm}),
characterized by a dilaton potential $V$
\beq\lb{dg}
A= \frac{1}{2} \int d^{2}x\sqrt{-g} \left (\phi R 
+\lambda^{2}V(\phi)\right),
\feq
where $\phi$ is a scalar (the dilaton) and $\la$ is parameter with dimensions of
a length$^{-1}$ (throughout this paper we will use natural 
units $c=\hbar=k_{B}=1$).
Two-dimensional dilaton gravity models have been used in various 
situations as effective description of 4D gravity . 
In particular, every spherically symmetric 4D solution  can be 
described in terms of a 2D dilaton gravity model.
The model (\ref{dg}) admits black hole solutions, which in the 
Schwarzschild gauge take the form \cite{Louis-Martinez:1993cc},
\beq\lb{bh}
ds^{2}= -
    \left( J(\lambda  r)- \frac{2M}{\lambda}\right)d\tau^{2}е 
    + \left( J(\lambda  r)- 
    \frac{2M}{\lambda}\right)^{-1}dr^{2},\quad \phi=\la r,
\feq   
where $M$ is the black hole mass and  $J=\int V d\phi$. The black 
hole horizon is located at $r=r_{h}$, with $J(\la r_{h})=2M/\la$.

We want to  find   acoustic analogs of the  2D black hole 
solutions (\ref{bh}). Obviously, the most natural candidate
is a fluid, whose motion is essentially one-dimensional.
Let us therefore consider the steady, locally irrotational  flow of a 
3D fluid which 
is barotropic and inviscid. If the transverse velocities (in the $z$ 
and $y$ directions) are small with respect to the velocity along the 
$x$ axis, we can set them to zero. The propagation of acoustic 
disturbances $\chi$ on the background fields  is determined, at the 
linearized level, by the wave equation 
$\nabla^{2}\chi=(1/\sqrt{-g})\partial_{\mu}(\sqrt{-g}
g^{\mu\nu}\partial_{\nu}\chi)=0$, where 
$g_{\mu\nu}$ is the so-called acoustic metric \cite{Visser:1997ux}.
In the case under consideration, the acoustic metric describes a 
one-dimensional slab geometry where the velocity is always along  the $x$ 
direction and the velocity profile depends only on $x$, 
\beq\lb{acoustic}
ds^{2}= \frac {\bar\rho_{0}}{c}\left[ -\left( c^{2} -v_{0}^{2}\right) dt^{2}
-2 v_{0} dxdt + dx^{2}\right ],
\feq
where $\bar\rho_{0}(x), v_{0}(x)$ are, respectively,  the background 
density and velocity of 
the fluid
and $c(x)= \sqrt{( dP/d\bar\rho_{0})}$ is the local speed of sound 
($P$ is the pressure).

The two metrics (\ref{bh}), (\ref{acoustic})  can be transformed 
one into the other using   a Painlev\'e Gullstrand-like 
coordinate transformation \cite{painleve,gullstrand} and identifying 
the function $J$ and the 
black hole mass $M$ in terms of the physical parameters  of the fluid.
It is convenient to work with a dimensionless  fluid density. 
This can be achieved by multiplying the metric (\ref{acoustic}) with 
a factor $\la^{-4}$ and defining the dimensionless density 
$\rho_{0}е=\la^{-4}\bar\rho_{0}е$.  The coordinate transformations and the 
identification relating  (\ref{bh}) with (\ref{acoustic})
are, 
\bea\lb{pg}\nonumber
r&=&\int \rho_{0}dx,\\  
\tau&=& t+ \int dx \frac{v_{0}}{ c^{2}-v_{0}^{2}},\\ \nonumber
J&=& \frac{2M}{\la} + \frac{\rho_{0}}{c}\left( c^{2} 
-v_{0}^{2}\right).
\eea
Notice that the coordinate $r$ (or equivalently the dilaton $\phi$) from the 
point of view of the 
fluid 
plays the role of an extensive  quantity. If the flux tube has a constant
section it is proportional to 
the mass of the fluid contained between some reference initial 
point and point $x$.
We stress the fact that the correspondence between 2D black holes (\ref{bh})
and the acoustic metric  
(\ref{acoustic}) is exact. This is to be compared with 
what happens for 4D static black holes, for which the correspondence 
with the acoustic metric can be established only up to a conformal 
transformation of the metric \cite{Visser:1997ux}.

Eqs. (\ref{pg}) allow to associate a one-dimensional slab geometry
(and a corresponding acoustic black hole) to every 2D black hole 
solution of the dilaton gravity model (\ref{dg}). They 
can be also used to construct acoustic analogs of 4D  spherically 
symmetric black hole solutions (e.g Schwarzschild). 
This is  not  surprising, being the causal structure of a 4D spherically 
symmetric  black hole completely encoded in its $(r,t)$ sections.

\section{2D Gravitational dynamics versus fluid dynamics}

In the previous section we have seen that given a generic 2D black 
hole we can always construct a 2D acoustic analog. However, the 
correspondence is purely kinematical. In order to formulate the 
correspondence at a dynamical level we need to compare the 
gravitational dynamics with the dynamics of the fluid, which determines
$ \rho_{0}, v_{0},c$.

At the dynamical level, the most striking difference between the two physical 
systems is that the gravitational dynamics has to be covariant under 
general coordinate transformations. In particular, this means that we 
have a huge redundancy of gauge degrees of freedom in the description 
of our 2D gravity system. These gauge degrees of freedom have to be 
fixed if we want to construct a  dynamical acoustic  analogue of the  
black hole. On the other hand, 2D dilaton gravity is purely topological.
We 
do not have any  physical propagating gravitational degree of freedom.
This leaves open the possibility that once the gauge 
freedom have been fixed, the dynamics for the resulting global 
gravitational degrees of freedom could match fluid dynamics.

The field equations for the 2D dilaton gravity  model (\ref{dg}) are
\beq\lb{fe}
R= -\la^{2} \frac{dV}{d\phi},\quad \nabla_{\mu}\nabla_{\nu}\phi -\frac{1}{2}
g_{\mu\nu}\la^{2}V=0.
\feq
We consider only static solutions and fix the diffeomorphisms 
invariance, 
choosing the Schwarzschild gauge for the 2D metric
\beq\lb{gf}
ds^{2}= -
    X(r) d\tau^{2}е 
    + X^{-1}(r) еdr^{2}.
\feq   
Using  Eq. (\ref{gf}) into the field equations (\ref{fe}), one finds 
after some manipulations that the  field equation for the dilaton $\phi$ 
are given simply by ${d\phi/dr}=\la$, which implies that the dilaton
is $\phi=\la r$. The field equations  for the global 
gravitational degree 
of  freedom $X(r)$  become instead
\beq\lb{fe1}
 \frac{d X}{dr}=\la V. 
\feq
One can also easily check that the black hole (\ref{bh}) is  solution 
of Eq. (\ref{fe1}).

The fundamental equations describing fluid dynamics are the Euler and 
the continuity equations.  The steady flow of the fluid giving rise to 
the one-dimensional slab geometry described in the previous section, 
is therefore governed by 
the equations:
\beq\lb{fd}
\bar\rho_{0}v_{0} \frac{dv_{0}}{dx} +\frac{dP}{dx} +\bar\rho_{0}е 
\frac{d\psi}{dx}=0, \quad \bar\rho_{0}(x)v_{0}(x) A(x)=const.,
\feq
where $P$ is the pressure, $\psi$ denotes  the  potential for  
external driving forces acting on the fluid (including Newtonian 
gravity) and  $A(x)$ is the area of the section 
of the flux tube.  We will  consider separately  the two cases:
a) The external potential $\psi$ is non-homogeneous and the section  
of the flux tube is constant. 
  b)  The external potential  is zero and  the section of the flux 
tube does depend  on $x$.

Let us first focus on case  a).
Using the barotropicity condition $P=P(\rho_{0})$, passing from 
the coordinate $x$ to the coordinate $r$ as defined in Eq. (\ref{pg}) 
and defining the new variables 
\bea\lb{nv}\nonumber
X&=& \frac{\rho_{0}}{c}\left( c^{2}- v_{0}^{2}\right),\\ 
Y&=& \rho_{0}c,\\ \nonumber
F&=& \ln \left(\frac{c}{\rho_{0}}\right),
\eea

the Euler and continuity equations (\ref{fd}) take, respectively, the form
\bea\lb{fd1}\nonumber
&&\frac{dX}{dr}=2\frac{dY}{dr}-X\frac{dF}{dr}+ 2e^{-F}\frac{d\psi}{dr}\\ 
&&\frac{d}{dr}\left[Y\left( Y- X\right)\right]=0.
\eea
One can now easily realize that the fluid dynamics equations 
(\ref{fd1}) can be made equivalent to the gravitational equations 
(\ref{fe1}) just by introducing the constraint
\beq\lb{constraint}
2\frac{dY}{dr}-X\frac{dF}{dr}+ 2e^{-F}\frac{d\psi}{dr}=\la V(\phi).
\feq
In fact, when constraint (\ref{constraint}) is enforced and the 
variable $X$ is identified  in terms of the gravitational variables 
as in Eq. (\ref{gf}) the Euler equation (\ref{fd1}) matches exactly 
Eq. (\ref{fe1}). 

Simple counting of the degrees of freedom and of the fundamental 
equations of the dynamics reveals that we can 
consistently map the fluid dynamics into the gravitational dynamics 
imposing constraints.  After gauge fixing, the gravitational system 
has only one degree of freedom, the function $X$ parametrizing the 
metric  in Eq. (\ref{gf}) and one equation. The scalar field $\phi$ 
is constrained to 
be proportional to $r$ and therefore represents just the spacelike
coordinate of the 2D spacetime. On the other hand, if the external 
potential is given, the fluid dynamics is 
characterized by three unknown functions $c,\rho_{0},v_{0}$ and two 
equations, so that we have the freedom of imposing an additional 
constraint. If the external potential $\psi$ has been fixed,
imposing the constraint (\ref{constraint}) is implicitly 
equivalent to   an equation of state $P=P(\rho_{0})$ for the 
fluid. From this point of view, the equation of state for the fluid 
may be seen as a consequence of the gravitational dynamics. This is
very similar to Jacobson's interpretation of the Einstein equation as 
an equation of state \cite{Jacobson:1995ab}.
Alternatively, if one first chooses a given equation of state
the constraint (\ref{constraint}) becomes an equation that determines
the external potential $\psi$.

Two-dimensional gravity allows us to define the mass $M$ of the black
hole solution (\ref{bh}) 
as a scalar, which on-shell becomes constant 
\cite{Mann:1992yv}
\beq\lb{mass}
M= \frac{1}{2\la}\left(\la^{2}\int V(\phi) d\phi - 
(\nabla\phi)^{2}\right).
\feq
Owing to the correspondence between the dynamics of 2D dilaton gravity 
and the previously discussed constrained fluid dynamics, we also 
expect the existence of a function $\Gamma$ of the fluid parameters $X,Y, 
F,\psi$ which is constant (independent of the flux tube coordinate 
$x$) by virtue of Eqs. (\ref{fd1}) and 
(\ref{constraint}).   This function is given by
\beq\lb{cons}
\Gamma= a \left[\int_{r_{0}}^{r} \left(-XdF+2 
e^{-F} d\psi\right)+2Y - X\right],
\feq
where $a$ and $r_{0}$ are arbitrary constants.
One can easily check that $d\Gamma/dr=0$ follows from Eq. (\ref{fd1})
and (\ref{constraint}). 
 
Eq. (\ref{cons}) allows us to find the acoustic counterpart of the 
black hole mass $M$.
Fixing the constant $a= \la/2$, using Eq. (\ref{constraint}) 
and reading from Eq. (\ref{pg}) the solution  for $X$, $X=J-2M/\la$,
one readily finds $M=\Gamma$. 

\section {Black hole thermodynamics  and fluid dynamics}

It is a well known fact that black hole thermodynamics follows 
directly from the gravitational field equations, once one has defined 
appropriately the thermodynamical parameters.
In the case of 2D gravity the thermodynamical state of the black hole 
is completely defined by 
temperature $T$,  energy (mass) $M$ and  entropy $S$.
The mass is given by Eq. (\ref{mass}).
The temperature can be defined in terms of the periodicity of the 
euclidean metric at the horizon, whereas the entropy  can be computed 
using Noether charge techniques \cite{Myers:1994sg} and turns out to be 
proportional to the dilaton evaluated at the black hole horizon. We have 
\beq\lb{ts}
T= \frac{\la}{4\pi} V(\phi_{h}),\quad   S=2\pi \phi_{h}.
\feq
The first principle of thermodynamics  follows directly from Eq. (\ref {fe1}).
In fact, using Eqs. (\ref{bh}) one has  $S= 2\pi 
J^{-1}(2M/\la)$, which in turn implies  $dM= 
(\la/4\pi) (dJ(r_{h})е/d\phi_{h}е) dS$. On  the other  hand Eq. 
(\ref{fe1})  implies  $dJ/d\phi=V$, which inserted in the 
previous equation and using Eq. (\ref{ts}) gives $dM=TdS$.

The correspondence between  gravitational and fluid dynamics 
described in the previous section allows us to define 
straightforwardly the thermodynamical parameters, $T_{a}, M_{a}, S_{a}е$ 
associated with the 
acoustic black hole. Moreover, the first principle of thermodynamics 
will become  consequence of the fluid dynamics  equations 
(\ref{fd1}),(\ref{constraint}). 

Although the phenomenon of Hawking radiation for acoustic black holes 
is a rather involved issue \cite{Unruh:2004zk}, one can still define 
also for the acoustic black hole a temperature  in  
terms  of  the  `` surface gravity''  or 
the periodicity of the Euclidean section at the acoustic 
horizon $c=v_{0}$. We have 
\beq\lb{ta}
T_{a}= \frac{1}{4\pi}\left(\frac{dX}{dr}\right)_{c=v_{0}е}= 
\frac{1}{4\pi }\left[\frac{1}{c}\frac{d}{dx}\left( 
c^{2}-v_{0}^{2}\right)\right]_{c=v_{0}}.
\feq
The above defined temperature coincides with the usual definition of 
temperature for acoustic black holes \cite{Visser:1997ux}.
Owing to the constraint (\ref{constraint}),  in our 2D case we can also 
write down an other, completely equivalent, expression for the 
temperature,
\beq\lb{ta1}
T_{a}= 
\left[\frac{1}{4\pi\rho_{0} }\left( 2\frac{dY}{dx}-X\frac{dF}{dx}+2e^{-F}
\frac{d\psi}{dx}\right)е
\right]_{c=v_{0}}.
\feq
For generic acoustic black holes it is not so easy to find a natural 
definition of mass. In the case under consideration we have a 
natural candidate for $M_{a}$, given by the quantity (\ref{cons}), 
which remains  constant along the flux  tube. 
Because it is a constant we can as well define $M_{a}$ as Eq. 
(\ref{cons}) evaluated at the acoustic horizon ,
\beq\lb{massac}
M_{a}е= \frac{\la}{2} \left[\int^{r_{h}е}_{r_{0}е}е dr\left( 
-X\frac{dF}{dr}+2 
e^{-F}\frac{d\psi}{dr}\right) +2Y(r_{h})\right]е,
\feq
where $r_{0}$ is chosen such that $M_{a}=M$.
The entropy of the acoustic black hole can be defined by analogy with 
the gravitational black hole, $S_{a}=2\pi \phi_{h}= 2\pi r_{h}$. 
Using the coordinate transformation (\ref{pg}) we get,
\beq\lb{entropy}
S_{a}= 2\pi \int^{r_{h}е}_{r=\infty} dx \rho_{0}(x),
\feq
where we have fixed the arbitrary integration constant in such way 
that the integration domain is the whole range of the coordinate $x$ 
outside the acoustic horizon. 
The entropy for the acoustic black hole is proportional to 
 the  total mass of the fluid outside the acoustic horizon. 
Notice that this  definition  
corresponds to the intuitive notion of entropy as an extensive quantity, 
which 
counts elementary degrees of freedom.

One can easily show,  differentiating Eq. (\ref{massac}) and using Eqs.
(\ref{ta1}) and (\ref{entropy}) that the thermodynamical parameters
satisfy the first principle $dM_{a}е=T_{a}еdS_{a}е$.

We can also show that the first principle can be derived directly from 
our constrained fluid dynamics. Evaluating the constraint  
(\ref{constraint}) at the acoustic horizon ($r=r_{h}$) 
 we get
$\la dJ(r_{h})=(2dY/dr-XdF/dr +2e^{-F}d\psi/dr)_{r_{h}} dr_{h}$.
Using Eqs. 
(\ref{ta1}) and (\ref{entropy}) together with $dJ(r_{h})=(2/\la) dM$,
we find $dM_{a}е=T_{a}е dS_{a}е$.

\section{ Flux tube with non-constant section}
Until now we have considered   a flux tube of constant 
section  with an non-homogeneous external potential.
The discussion can be  easily extended to a flux tube with 
non-constant section. In this latter case the  presence of a 
external potential is an unnecessary complication.
For this reason we only consider the dynamics of a fluid with 
zero external potential and a non-constant section.
In this case the fluid dynamics equations (\ref{fd}), written in terms 
of the variables  defined in  Eq. (\ref{nv}) are
\beq\lb {fd2}
\frac{dX}{dr}=2\frac{dY} {dr}-X\frac{dF}{dr},\quad 
\frac{d}{dr}\left[\left(Y-X\right)YA^{2}\right]=0.
\feq
Introducing the constraint 
\beq\lb{con2}
2\frac{dY} {dr}-X\frac{dF}{dr}=\la V,
\feq
Eqs. (\ref{fd2}) can be made equivalent to the
gravitational equations (\ref{fe1}).

The function that is constant owing to Eqs. (\ref{fd2}), (\ref{con2}) 
and can be 
therefore interpreted as the black hole mass is now
\beq\lb{cons1}
\gamma= a \left[\int^{r}_{r_{0}е}е \left(-XdF\right)+2Y- 
X\right].
\feq
Also in this case, once $A(x)$ has been fixed, the constraint 
 (\ref{con2}) is equivalent  to an equation of state for the fluid.
On the other hand if the equation of state of the fluid is given, 
Eq. (\ref{con2}) determines $A(x)$. In this latter case, Eq.
(\ref{con2}) represents a 
``geometrical constraint'' for the fluid flow  of the same kind of 
that 
discussed in Ref. \cite{Barcelo:2001ca} in relation to the
realizability of acoustic horizons.

The mass  and  temperature of the acoustic black hole are now
\beq\lb{ap}
M_{a}е= \frac{\la}{2}\left[ \int^{r_{h}е}_{r_{0}е}ее dr\left( 
-X\frac{dF}{dr}\right)+2Y(r_{h}\right]е,\quad
T_{a}=\left[\frac{1}{4\pi\rho_{0} }\left( 2\frac{dY}{dx}-X\frac{dF}{dx}\right)е
\right]_{c=v_{0}},
\feq
whereas the entropy is always given by Eq. (\ref{entropy}).
One can easily check the validity  of  the first principle of 
thermodynamics and that this principle is a consequence of the 
equations of the the fluid dynamics (\ref{fd2}), (\ref{con2}).

\section{Solutions of the constrained fluid dynamics}
Apart from its intrinsic theoretical interest, our approach could 
also have a direct experimental realization. Given a 
four-dimensional, spherically symmetric,  gravitational black hole,
one  can derive by 
dimensional reduction the corresponding 2D dilaton gravity model 
(\ref{dg}) and related 2D black hole (\ref{bh}). The 2D black hole 
solution is completely characterized by the function $J$ and the mass 
$M$. Once  $J(r)$ and $M$ are given,  we can construct 
explicitly the 
corresponding 2D acoustic black hole and, at least in principle, 
realize the experimental set-up, which will produce the acoustic 
black hole. 

In order to have a well-defined physical system, we need  to solve 
the fluid dynamics  equations, i.e we need to know $\rho_{0}, c ,v_{0}$ as
function of the  flux tube coordinate $x$.
Equations (\ref{fe1}) and the second equation in (\ref{fd1}) (or in 
(\ref{fd2})) can be easily integrated 

\beq\lb{e4}
X=J-\frac{2M}{\la},\quad 
Y=\frac{1}{2}\left(X+\sqrt{X^{2}+\alpha^{2}}\right),
\feq 
where $\alpha$ is a constant in case a), whereas $\alpha=const/A(x)$ 
in case b). 

To solve the 
dynamics completely, the constraint (\ref{constraint}) (or \ref{con2}) has to be 
enforced. This can be done in two nonequivalent ways. $1.$ One can 
choose a given barotropic equation of state for the fluid and the 
constraint (\ref{constraint}) ((\ref{con2})) determines the external 
potential $\psi$ (the profile $A(x)$ for the flux tube). $2.$ The  
external potential $\psi$ (the flux tube profile $A(x)$) is given and 
the constraint (\ref{constraint}) ((\ref{con2})) determines the 
barotropic equation of state for the fluid.
Because one usually operates with a fluid with a given equation of 
state, the physically relevant situation is obviously the first one. 
Therefore we will consider here only  possibility $1.$.
To handle with a definite and common situation, we will consider a 
fluid with a constant speed of sound $c$. The barotropic equation of 
state is $P=c^{2}\rho_{0}$, which expressed in terms of the variables
(\ref{nv}) gives
\beq\lb{e10}
Ye^{F}=c^{2}.
\feq

Let us first consider the case of flux tube of constant section (case 
a)). Inserting Eqs. (\ref{e10}) and (\ref{e4}) into the constraint 
(\ref{constraint}) and using Eqs. (\ref{nv}),(\ref{e4}),
(\ref{e10}) one  
  obtains  the solution for $v_{0}(X), \rho_{0}(X)$ and $\psi(X)$,
\beq\lb{e11}
v_{0}= \frac{\alpha c}{2Y},\quad \rho_{0}= \frac{1}{c} Y,\quad 
\psi(X)= c^{2}\left(\frac{X}{2Y}-\ln Y\right),
\feq
where $Y(X)$ is given as in Eq. (\ref{e4}).
The acoustic horizon is located at $X=0$ ($Y=\alpha/2$). At the 
horizon the density remains finite, $\rho_{0}=\alpha/2c$.
This is the best we can do in the general case. The calculations can 
be completed once a specific 2D gravitational model ( a specific 
function $J$) is  given.  Taking into account  that $\rho_{0}=dr/dx$, 
the previous equation for $\rho_{0}$ becomes a differential equation, 
which determines the coordinate transformation $r=r(x)$.

Let us now consider the case of a  flux tube on non-constant section 
(our case  b)).  We can use  Eqs. (\ref{e4}) to obtain
\beq\lb{e12}
A^{2}\propto\frac{1}{Y(Y-X)}.
\feq
Notice that $Y\ge X$.
In order to  determine $Y(X)$  we   use Eq. (\ref{e10}) in the constraint
(\ref{con2}) to get
\beq\lb{e13}
2\frac{dY}{dX} +\frac{X}{Y} \frac{dY}{dX}=1.
\feq

The solution of this equation is given in implicit form by 
\beq\lb{e14}
X=2Y\ln Y,
\feq
where for simplicity we have set to zero the integration constant.
We can now write the solution of our   fluid dynamics equation in the form
\beq\lb{e15}
v_{0}= cе\sqrt{1- \frac{X}{Y}},\quad \rho_{0}=
\frac{Y}{c},
\feq
where $Y(X)$ is implicitly defined by Eq. (\ref{e14}).
The horizon is located at $X=0$, the subsonic region for $X>0$.
From very general arguments (see for instance Ref. \cite{Visser:2001fe})
one expects that the acoustic  horizon forms  at a minimum of the
area of the flux tube section ( the so called Laval nozzle). 
Let us show that this is also true in our case.
From Eqs. (\ref{e12}) and (\ref{e13})  one finds
\beq\lb{e15a}
A'\propto \frac{\left(1-2Y'\right)}{Y-X},
\feq
where the prime denotes derivation with respect to $X$.
On the other hand,  Eq. (\ref{e13}) tells us that at the horizon 
we have $Y'=1/2$ and that in the subsonic (supersonic) region
$Y'<1/2$ ($Y'>1/2$). It follows 
immediately that $A'=0$ at the horizon and that $A'>0$ ($A'<0$) in the 
subsonic (supersonic) region, i.e that in the  horizon region the flux tube must   
take the form of a Laval nozzle.

\section{Two examples: 2D Anti-de Sitter and 4D Schwarzschild 
black hole}

In this section  we will consider two particular cases 
as illustration of our general 
procedure . 
Our first example is the 2D anti-de Sitter (AdS) black hole 
\cite{Cadoni:1994uf}. The  corresponding 2D dilaton gravity model (\ref{dg})
has a constant
potential $V(\phi)=2$, from which follows $J(r)=\la^{2}r^{2}$.
The black hole solution (\ref{bh}) describes a 2D AdS spacetime with an 
horizon at $r=\sqrt{2M/\la^{3}}$. The corresponding acoustic black 
hole  can be easily constructed using our general procedure. Let us 
consider separately the two cases of constant and non-constant 
flux  tube section.

\subsection{Flux tube with constant section}

\begin{figure}[h]
\includegraphics[angle=0.0,width=7.5cm]{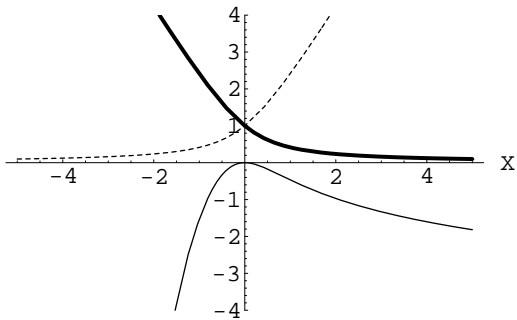}
\caption{\label{frame}
Behavior of the normalized  parameters $\psi/c^{2}$ (normal line), 
$v_{0}/v_{0}(r_{h})е$(bold line) and $\rho_{0}/\rho_{0}(r_{h})е$
(dashed line) as a function of the variable $X$ for a flux tube  with
constant section and non-homogeneous external potential characterized 
by  $\alpha=1$. 
The  velocity and fluid density are normalized to
their horizon values. The acoustic horizon is located  at $X=0$}
\end{figure}

Considering a flux tube  in presence of a non-homogeneous  
external potential and with constant section, the  acoustic analog
of the 2D AdS black hole can be easily found inserting $X= \la^{2}r^{2}- 2M/\la$ 
into Eqs. (\ref{e11}).
Using the expression for $\rho_{0}$ given in equation (\ref{e11}) 
one can also find  
the relation between the radial coordinate $r$
of 
the gravitational black hole and the spatial coordinate $x$ of the 
acoustic black hole.
The $0\le r<r_{h}$ interior of the AdS black hole is described by the supersonic
region of the acoustic black hole. At $r=0$ the fluid parameters 
$\rho_{0},v_{0}$ and the external potential are finite. For $r>0$ 
$v_{0}$ decreases  and $\rho_{0}$ grows monotonically. Also the
external potential $\psi$ grows for $r>0$ but reaches its maximum at 
the horizon $r=r_{h}$. 
At the acoustic horizon, $v_{0}=c$ and
both the density $\rho_{0}$ and the external potential remain finite.
The asymptotic  region 
$r\to\infty$ of the gravitational black hole (the timelike conformal
boundary of the 2D AdS spacetime) corresponds to the 
subsonic region 
 of the acoustic black hole. 
Going in this region $v_{0}\to 0$ ,   the fluid  density $\rho_{0}$ 
and fluid pressure $P$ diverge 
as expected owing to the continuity equation. 
Conversely, the potential  $\psi$ of the external force acting on the fluid
diverges logarithmically as $-\ln r$. 
The behavior of the parameters $\psi, v_{0}, \rho_{0}$, normalized to their horizon 
values  and for $\alpha=1$ are shown if Fig. (\ref{frame}) as a function of the variable
$X=\la^{2}r^{2}- 2M/\la$.

\subsection{Flux tube with non-constant section}

If we use  a  flux tube of non-constant section to mimic the 2D AdS
black 
hole, we get the typical shape of a converging/diverging Laval nozzle. 
We can find the 
solution for the fluid equations using  $X= \la^{2}r^{2}- 2M/\la$ into 
Eqs. (\ref{e12}), (\ref{e14}) (\ref{e15}). 
In this case the fluid can 
be used to describe only the horizon region  of the AdS black hole, but not 
the full spacetime. In fact, if $A^{2}$ of Eq. (\ref{e12}) has to 
stay positive $X$ and because of Eq. (\ref{e14}) it
follows  that 
$-2/e\le X\le \sqrt{e}$ 
and correspondingly   $1/e\le Y\le \sqrt{e}$.
The supersonic region of the acoustic black hole describes the region 
$\sqrt{2M/\la^{3}-2/(e\la^{2})}=r_{m}\le r< r_{h}=\sqrt{2M/\la^{3}}е$
of the AdS black hole interior.
Analogously, the subsonic region  can be used  to describe  only the region 
$ \sqrt{2M/\la^{3}}=r_{h}е< r\le r_{M}=
\sqrt{2M/\la^{3}+\sqrt{e}/\la^{2}}$ of the AdS black hole.

\begin{figure}[h]
\includegraphics[angle=0.0,width=7.5cm]{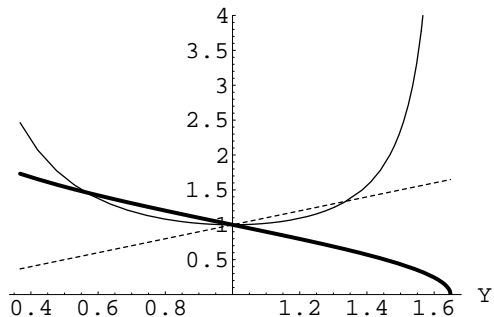}
\caption{\label{frame1}
Behavior of the normalized  parameters $A/A(r_{h})$ (normal line), 
$v_{0}/v_{0}(r_{h})е$(bold line) and $\rho_{0}/\rho_{0}(r_{h})$
(dashed line) as a function of the variable $Y$ for a flux  tube with
non-constant section. 
The flux tube section, the velocity and fluid density are normalized to
their horizon values. The acoustic horizon is located  at $Y=1$}
\end{figure}

The profile of the flux tube is that of the Laval nozzle  depicted in 
Fig. (\ref{frame1}), which also shows the behavior of the fluid velocity and
fluid density, normalized to their horizon values, as a function of
the variable $Y$ defined in Eq. (\ref{e14}). 
The  flux tube section $A$ starts at a finite value at $r=r_{m}$
decreases in the supersonic region of the acoustic black hole till it 
reaches its minimum   $A(r_{h})$ at the
  horizon. 
In the subsonic 
region $r_{h}<r<r_{M}$,  $A$ grows and diverges 
at $r=r_{M}$. The fluid velocity decreases monotonically  starting from
$v_{0}(r_{m})=\sqrt{3}c$ in the supersonic region to reach first its
horizon value $v_{0}(r_{h})=c$ and then  $v_{0}(r_{M})=0$,
as expected owing to the equation of continuity.
Finally, the fluid  density $\rho_{0}$, grows monotonically from 
 $\rho_{0}(r_{m})= (ec)^{-1}е$ in the supersonic
region to the horizon  value  $\rho_{0}(r_{h)}е=c^{-1}$ to reach
$\rho_{0}(r_{M})= \sqrt{e}c^{-1}е$ in the subsonic region.

Our second example is the 4D Schwarzschild black hole. By means of 
a dimensional reduction,  retaining 
only the radial modes, 4D Einstein gravity can be  described by the 2D
gravity model (\ref{dg}) with dilaton potential 
$V(\phi)=1/\sqrt{2\phi}$ and with  $\la^{-2}$ identified as the 4D Newton 
constant $G$. The 2D black hole solution (\ref{bh}) becomes, 
\beq\lb{sc}
ds^{2}= \left(\sqrt{2\la r}- \frac{2M}{\la}\right)d\tau^{2}+
\left(\sqrt{2\la r}- \frac{2M}{\la}\right)^{-1}еdr^{2}.
\feq
This solution describes the 2D sections of the 4D Schwarzschild 
solution $ds^{2}=-(1- 2MG/R)d\tau^{2}+(1- 2MG/R)^{-1}dR^{2}$.
In fact the metric (\ref{sc}) can be put in the Schwarzschild form
by means of the Weyl rescaling of the 2D metric $g_{\mu\nu}\to (1/
\sqrt{2\phi})g_{\mu\nu}$ and  the coordinate transformation 
$r=(\la/2) R^{2}е$. 
Mass, temperature and entropy of the 2D black hole are invariant under 
Weyl transformations of the metric \cite{Cadoni:1996bn}. In fact the temperature and 
entropy of  the 2D black hole (\ref{sc}), given by $T= \la^{2}/8\pi M,\,
S= (4\pi M^{2}/\la^{2})$, match exactly those of the 4D Schwarzschild 
black hole after setting $\la^{-2}=G$.

The acoustic analogue of the Schwarzschild black hole can be 
constructed, working with a flux tube  of constant section  
putting $X= \sqrt {2\la r} -2M/\la=\la R -2M/\la$  into 
Eqs. (\ref{e11}),
 (or into Eqs. (\ref{e12}), (\ref{e14}), (\ref{e15}) if 
working with a  flux tube  of non constant section).

The features of the acoustic Schwarzschild black hole are 
qualitatively similar to 
those of the previously discussed AdS acoustic black hole.

\section{Summary  and outlook}
In this paper we have developed a 
general method for  constructing   acoustic analogs of 2D black hole 
solutions.  The procedure can be also used  for 4D black holes that 
admit an effective 2D description. In particular this is the case of
spherically symmetric 4D black holes.  
We have also shown that after fixing the gauge 
degrees of freedom, the 2D gravitational 
dynamics is equivalent to a constrained fluid dynamics.
The correspondence between  gravitational and fluid dynamics has allowed a 
natural definition of the thermodynamical parameters mass, temperature 
and entropy of the  acoustic black hole. 
Moreover, we have shown that   the first principle of 
thermodynamics follows from the constrained fluid dynamics.

The results presented in this paper  represent just a first step 
for finding condensed matter  systems that mimic  gravitational 
dynamics.
It is not hard to identify the two main (and related) limitations 
of our approach.
First, our discussion holds for static gravitational configurations.
In many interesting situations, for instance the computation of the 
back-reaction for an evaporating black hole, one has to deal with non 
static solutions.
The generalization of our method  to the non static case may be 
extremely non trivial, in particular the fixing of  the gravitational 
gauge degrees of freedom necessary to reproduce the fluid dynamics.
The  simplicity of 2D approach can be very useful also for  
treating  the non static case.
Second, in this paper we have considered a 2D gravitational model 
without matter. The absence of sources for the gravitational field
represent a strong simplification for the dynamical equations. 
On the other hand, we need the presence of sources if we want to 
describe a realistic situation.
Finding a fluid dynamic counterpart of  the matter stress-energy 
tensor appearing in the gravitational field equations may
not be a  easy task.

From a more technical point of view, if we work with a fluid with a 
given equation of state 
the practical realization of a physical system that mimics  the 2D 
black hole requires a particular careful choice of an external 
potential (flux tube of constant section) or of the flux tube profile
(flux tube of nonconstant section). 
\begin{acknowledgments}
I am grateful to A. Fabbri, S. Liberati a S. Mignemi for helpful
comments.
\end{acknowledgments}


\end{document}